\documentclass{PoS}
\usepackage{amsmath}
\input myunits.sty
\input mymath.sty
\usepackage{subfig}
\usepackage[sort&compress,square,comma,numbers,merge]{natbib}
\graphicspath{{./fig/}}
\DeclareGraphicsExtensions{.png,.pdf,.jpg,.mps}
\usepackage{grffile}

\title{Precision Studies of Light Mesons at COMPASS}

\ShortTitle{Light Mesons at COMPASS}

\author{\speaker{Bernhard Ketzer}\thanks{Present address:
    Helmholtz-Institut f\"ur Strahlen- und Kernphysik, Universit\"at
    Bonn, D-53115 Bonn, Germany.} for the COMPASS Collaboration\\
        Physik Department, Technische Universit\"at M\"unchen,
        D-85748 Garching, Germany\\
        E-mail: \email{Bernhard.Ketzer@cern.ch}}

\abstract{%
The COMPASS experiment at CERN's SPS investigates the structure and
excitations of strongly interacting systems. Using reactions of
$190\,\GeV/c$ pions with protons and nuclear targets, mediated by 
the strong and electromagnetic interaction, 
an unprecedented
statistical precision has been reached allowing new insight into the
properties of light mesons. For the first time the diffractively
produced $3\pi$ final state has been analyzed simultaneously in bins
of invariant mass and four-momentum transfer using a large set of 88
waves up to a total angular momentum of 6. In addition to a precise
determination of the properties of known resonances and including a
model-indepedent analysis of the $\pi\pi$ $S$-wave isobar, a new
narrow axial-vector state coupling strongly to $f_0(980)$ has been
found in previously unchartered territory. By selecting reactions with
very small four-momentum transfer COMPASS is able to study processes
involving the exchange of quasi-real photons. These provide clean
access to 
low-energy
quantities such as radiative couplings and polarizabilities of mesons, 
and thus constitute a test of model 
predictions such as chiral perturbation theory.  
}

\FullConference{XV International Conference on Hadron Spectroscopy-Hadron 2013\\
		4-8 November 2013\\
		Nara, Japan}

\begin{document}

\section{Introduction}
\label{sec:intro}
Despite more than fifty years of research, the light-meson excitation
spectrum is still far from being understood. In the original quark
model proposed in 1964 
light
mesons are composed of quark-antiquark ($q\overline{q}$) pairs with
$LS$-coupling of total spin $S$ and orbital angular momentum $L$ between the
quarks. This static model yields a surprisingly accurate description
and classification of many of the light mesons known today
\cite{Beringer:2012zz}. 
In this model, the pseudoscalar and vector mesons are the $L=0$
combinations with total spin $S=0$ and $S=1$, respectively, yielding a
total angular momentum $J=0$ or $1$. Parity and charge-conjugation
parity\footnote{for states composed of a quark and its own antiquark}
are given by $P=(-1)^{L+1}$ and $C=(-1)^{L+S}$, respectively.  
The classification of light mesons with scalar quantum numbers
$J^{PC}=0^{++}$ is more difficult. 
One possible interpretation is that the scalars with masses below
$1\,\GeV/c^2$ form a nonet with an inverted mass hierarchy.   
The high masses of the
$a_0(980)$ and the $f_0(980)$ and their large coupling to $K\bar{K}$
could be explained by interpreting them as tightly bound tetraquark
states \cite{Jaffe:1976ig} or $K\bar{K}$ molecule-like objects
\cite{Weinstein:1983gd}. 
Other interpretations favor an ordinary
$q\bar{q}$ nonet consisting of $f_0(980)$, $a_0(980)$,
$K_0^\ast(1430)$, and $f_0(1500)$\ \cite{Minkowski:1998mf,Ochs:2013gi}. 

Quantum Chromodynamics (QCD)  
provided a justification for the $q\overline{q}$ rule as a possible
singlet representation\footnote{together with the $qqq$ system for
  baryons} 
of the underlying fundamental color $SU(3)$ symmetry.   
Being composed of a strongly coupled 
system of highly relativistic light quarks $u$, $d$, $s$, however,
calculations of the properties of hadrons using perturbative QCD are
bound to fail. Furthermore, QCD allows the existence of whole new
classes of mesons, including systems being composed of 4 quarks
($q\overline{q}$ molecules or tetraquark states) and those where gluonic
degrees of freedom contribute to the quantum numbers (hybrids,
glueballs). Despite a long history of experiments only some isolated
candidates for such states were identified (see
\cite{Klempt:2007cp,Crede:2008vw,Ketzer:2012vn} for recent reviews).  
However, the phenomenological
picture is still very much dominated by the nonets of the $q\overline{q}$
model. Recently, 
first ab-initio calculations of hadron properties have been performed
by numerically solving QCD  
on a Euclidean space-time lattice
\cite{Aoki:2008sm,Durr:2008zz,Dudek:2010wm}.  
Similarly to the
experimental situation, 
the quark-model states seem to be confirmed by
the lattice simulations; some recent calculations also predict a full
set of multiplets of non-$q\overline{q}$ states, e.g. hybrids
\cite{Dudek:2011bn}. 

\section{COMPASS Detector}
\label{sec:compass}
The COMPASS experiment at the CERN SPS set out to solve some of the
open questions  
in light-meson spectroscopy by increasing the world data sample
for exclusive 
reactions of high-energy ($190\,\GeV/c$) hadrons with various targets
by a factor of ten to hundred. 
In COMPASS kinematics, three different mechanisms contribute to the
production of 
a system $X$:  
diffractive  
dissociation and central production, 
which can be described to proceed via the exchange of one or two
Reggeons $\mathbb{R}$, respectively, between the beam hadron and the target
particle $N$, and photo-production in the Coulomb field of a nucleus
(Primakoff reactions) at
very low values of momentum transfer.   
In order to detect the final-state particles from these processes over
a wide angular range with excellent resolution,
COMPASS is built as a two-stage magnetic spectrometer equipped with 
tracking, particle identification and calorimetry, providing a very
uniform acceptance for  
neutral and charged particles over a broad kinematical range
\cite{Abbon:2007pq}. The 
incoming beam particles ($\pi$, $p$, $K$) are identified by a pair of
CEDAR detectors. For the results reported in this paper a negative
beam consisting of $96.8\%$ $\pi^-$ was used. 
For the diffractive measurements
a $40\,\Cm$ long liquid-hydrogen target was employed, while for the
Primakoff reactions a solid target disk made of either lead
or nickel was installed. The target is surrounded by 
a Recoil Proton Detector (RPD) consisting of two concentric layers of
scintillator bars. Events containing particles emerging at angles
larger than the acceptance of the spectrometer are vetoed by a
dedicated sandwich counter. The information of several trigger
detectors is used in order to select exclusive events according to the
physics case. 

\section{Diffractive Dissociation Reactions}
\label{sec:diff}

\subsection{Kinematics}
\label{sec:diff.kin}
\begin{figure}[tbp]
  \centering
  \includegraphics[width=0.32\columnwidth]{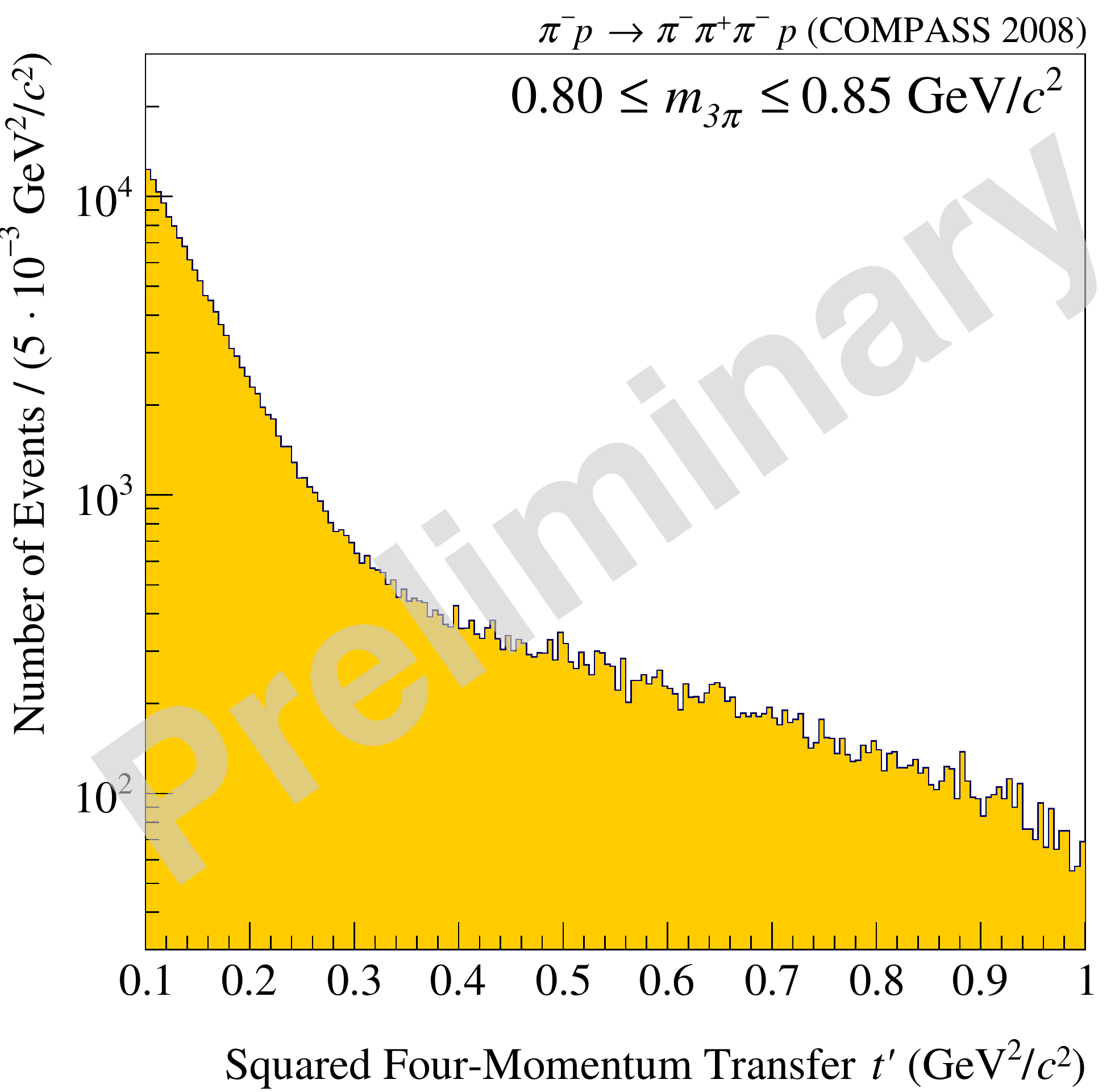}
  \hfill
  \includegraphics[width=0.32\columnwidth]{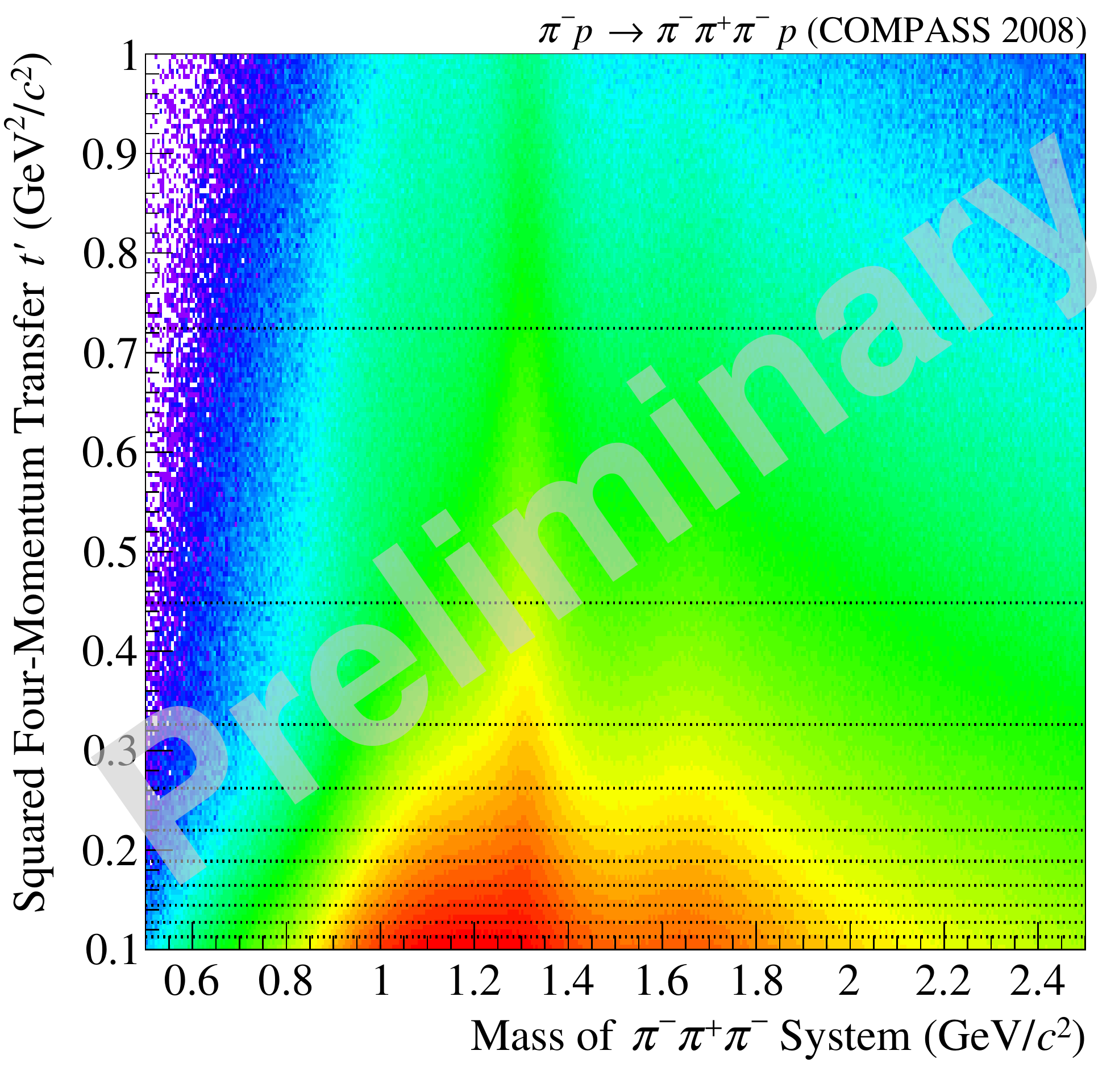}
  \hfill
  \includegraphics[width=0.32\columnwidth]{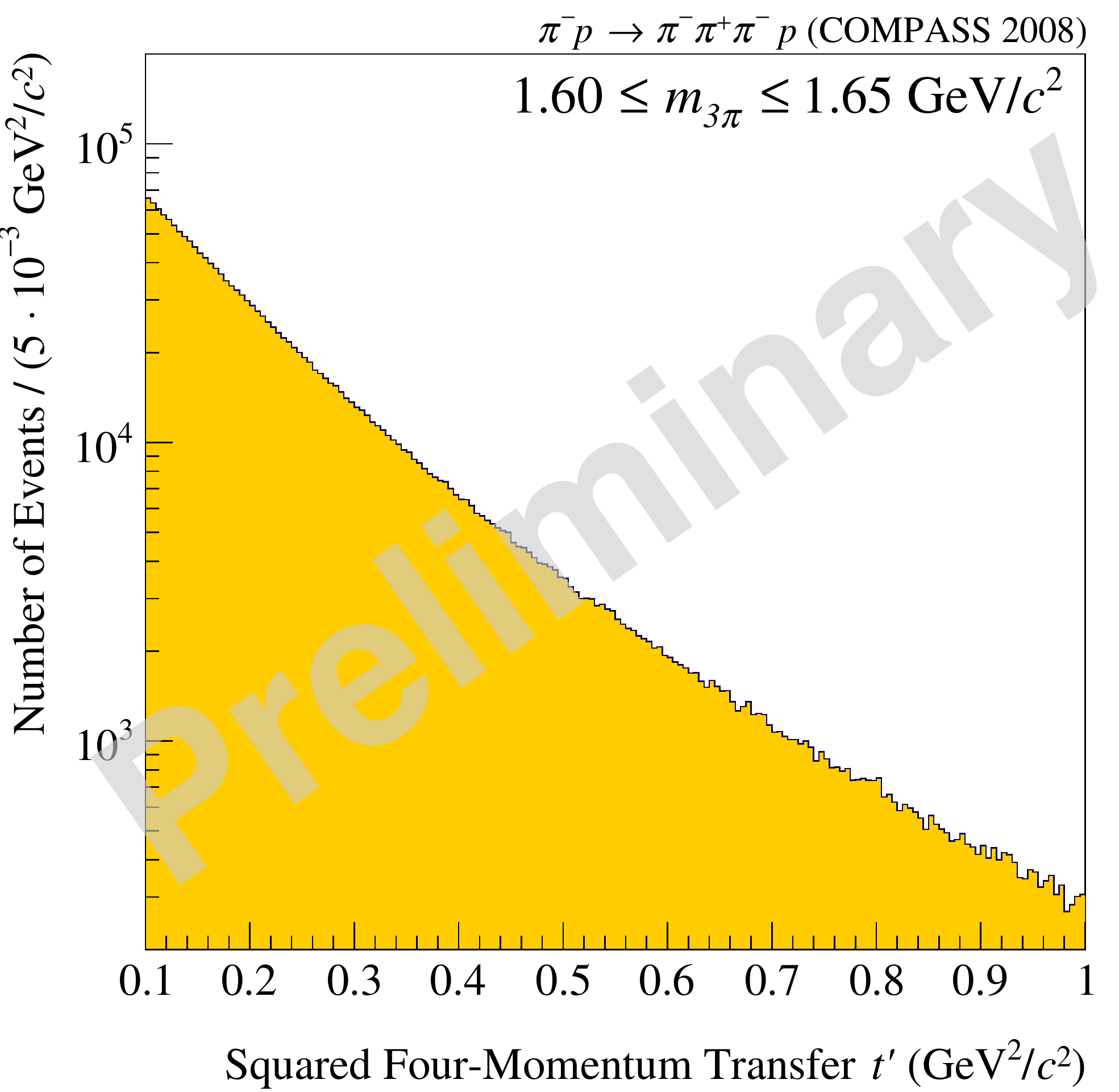}
  \caption{$3\pi$ final state, (center) distribution of events as a
    function of 
    $3\pi$ invariant mass $m_{3\pi}$ and 4-momentum transfer $t'$,
    (left) $t'$-distribution for $0.80\leq m_{3\pi}\leq
    0.85\,\GeV/c^2$, (right) $t'$-distribution for $1.60\leq m_{3\pi}\leq
    1.65\,\GeV/c^2$.} 
  \label{fig:pi-pi-pi+.m-vs-t}
\end{figure}

For the study of diffractive dissociation reactions 
the trigger 
requires an incoming beam
particle hitting the target and a signal from a recoil proton in the
RPD, resulting in a 4-momentum transfer
$t<-0.07\,\GeV^2/c^2$. Exclusive events are selected by energy
conservation between incoming and outgoing particles and by transverse
momentum balance between incoming and outgoing particles including the
recoil proton.  
Here we focus on the final state containing three charged pions,
$\pi^-\pi^-\pi^+$, with about $5\EE{7}$ events after the
above-mentioned cuts, covering invariant masses up to
$3\,\GeV/c^2$. The corresponding analysis for the $\pi^-\pi^0\pi^0$
final state is covered elsewhere in these proceedings. 
Figure~\ref{fig:pi-pi-pi+.m-vs-t} (center) displays the 
distribution of events as a function of the 
invariant mass $m_{3\pi}$ 
of the three-pion system and $t'$,
defined as the reduced squared 4-momentum transfer 
$\vert t \vert - \vert t \vert_\mathrm{min}$
to the recoiling target nucleon beyond the kinematic
minimum $\vert t
\vert_\mathrm{min}$. Figures~\ref{fig:pi-pi-pi+.m-vs-t} (left) and
(right) show the projection onto the $t'$ axis for two different bins
of $m_{3\pi}$, respectively. 
Evidently there is a strong correlation between these two variables,
illustrating the need to analyze the data not only in bins of $m_{3\pi}$
but also in bins of $t'$. 

\subsection{Partial-Wave Analysis}
\label{sec:diff.indep}
To disentangle the resonances contributing to a given final state a
partial-wave analysis (PWA) is performed  
using the phenomenological approach
of the isobar model. In this model the production and the
decay of a state $X$ with quantum numbers $J^{PC}M^\epsilon$
factorize. 
Here, $M$ is the modulus of the spin projection onto a
quantization axis, and $\epsilon$ the reflectivity
\cite{Chung:1974fq}. The decay is
described as a series of   
sequential two-body decays into intermediate resonances (isobars),
which eventually decay into the final state observed in the
experiment. 
A partial wave is fully characterized by the quantum numbers of $X$,
the isobars, their total spin and the orbital angular momentum between
them. In the case of the three-pion final state, we allow for all
di-pion isobars established by the PDG \cite{Beringer:2012zz} up to a
mass of $1.7\,\GeV/c^2$:
$(\pi\pi)_S$ wave (containing the broad $f_0(500)$ and the
$f_0(1370)$) \cite{Au:1986vs}, $f_0(980)$, $f_0(1500)$, $f_2(1270)$, 
$\rho(770)$, $\rho_3(1690)$. The analysis makes use of an event-based
extended log-likelihood fit in
$20\,\MeV/c^2$ wide bins of $m_{3\pi}$ and in 11 bins of $t'$, chosen
to equalize the number of events in each bin and 
indicated by horizontal lines in Fig.~\ref{fig:pi-pi-pi+.m-vs-t}
(center). The fit includes the largest wave set ever used in such an
analysis: 80 waves with positive reflectivity and spin up to 6, seven
waves with 
negative reflectivity, and a flat background wave representing
three-pion phase space. Full coherence is assumed for waves with
positive and negative 
reflectivity, respectively, while the set of positive- and
negative-reflectivity 
waves and the flat wave are added incoherently. Only waves with negligible
population have been omitted from the fit.   

\subsection{Fits of Spin-Density Matrix}
\label{sec:diff.mdep}
\begin{figure}[tbp]
  \centering
  \includegraphics[width=0.85\columnwidth]{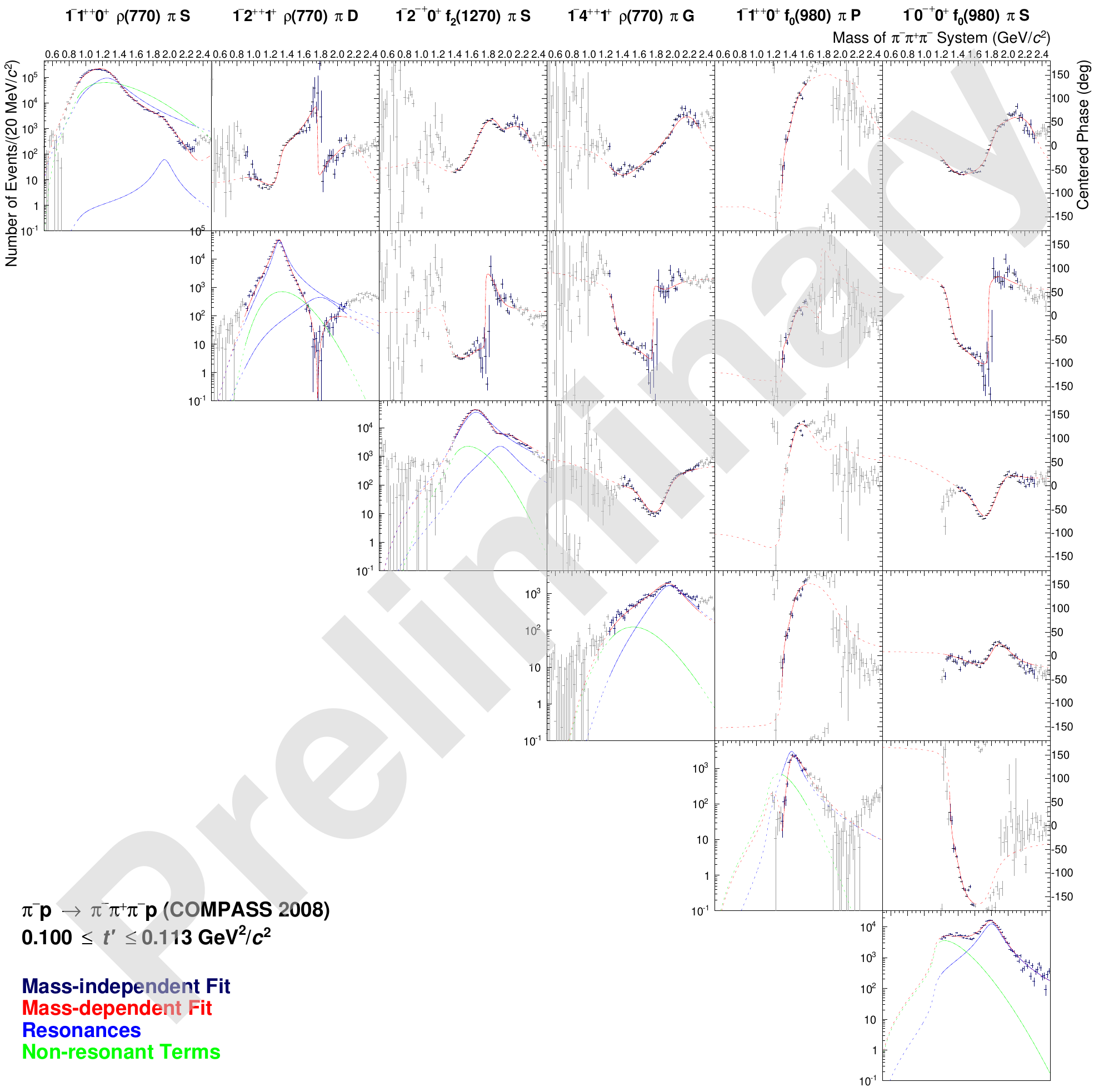}
  \caption{Mass and $t'$-dependent fit of the spin-density matrices
    for six selected 
    waves (columns labeled at the top, rows analogously), shown here
    for $0.100\leq t'\leq 
    0.113\,\GeV^2/c^2$. The plots along 
    the diagonal correspond to the intensities in the respective
    waves, while the off-diagonal plots show the phase differences
    between the wave corresponding to the column and the one
    corresponding to the row, respectively. The black data points are
    the result of the PWA in mass and $t'$ bins, the curves indicate
    the fit model.} 
  \label{fig:pi-pi-p+.rho_matrix}
\end{figure}

The result of the PWA described in Section~\ref{sec:diff.indep} is one
independent spin-density matrix for each
$m_{3\pi}$ and $t'$ bin, containing all waves used in this particular
bin. In a second step, a model is applied in a
$\chi^2$ fit to  
describe the mass and $t'$-dependence of these
matrices, where for computational reasons only a few waves, in our
case six, are considered. For each wave the model includes 
resonant contributions, usually parametrized in terms of relativistic
Breit-Wigner functions with dynamic widths and parameters
\emph{independent} of $t'$, 
and non-resonant 
contributions added coherently, in our case parametrized by empirical functions
$\exp[-p(t')q^2]$, with $p(t')$ a polynomial of the reduced 4-momentum
transfer and $q$ the break-up momentum for 2-body decay.  
Figure~\ref{fig:pi-pi-p+.rho_matrix} displays the 
result of such a fit to the spin-density matrices of the first step,
displayed here as a function of $m_{3\pi}$ and for the smallest $t'$
bin. 

Apart from the observation of known resonances with very small statistical
uncertainty, the most important features are (i) the need for a second
$a_1$ resonance (top row), (ii) the need for a 
second $a_2$ state interfering destructively with the $a_2(1320)$
(second row/column), both at masses close to $2\,\GeV/c^2$, and (iii)
the observation of a clear peak of the
$1^{++}0^+\,f_0(980)\,\pi\,P$ wave intensity at a mass of
$1.42\,\GeV/c^2$ combined 
with a phase motion close to $180^\circ$ with respect to all other
waves (5th row/column). In our model, this previously unobserved 
object with axial-vector quantum numbers, coupling exclusively to
$f_0(980)\pi$, is well described by a 
Breit-Wigner resonance, tentatively called 
$a_1(1420)$, with a rather small width of about $140\,\MeV/c^2$. 
The interpretation of this structure as a genuine resonance or,
due to its proximity to the $K^\ast\bar{K}$ threshold, as a dynamic
effect \cite{Basdevant:1977ya} including rescattering of the $K\bar{K}$ to
$f_0(980)$, is still to be clarified. 

\begin{figure}[tbp]
  \centering
  \includegraphics[width=0.32\columnwidth]{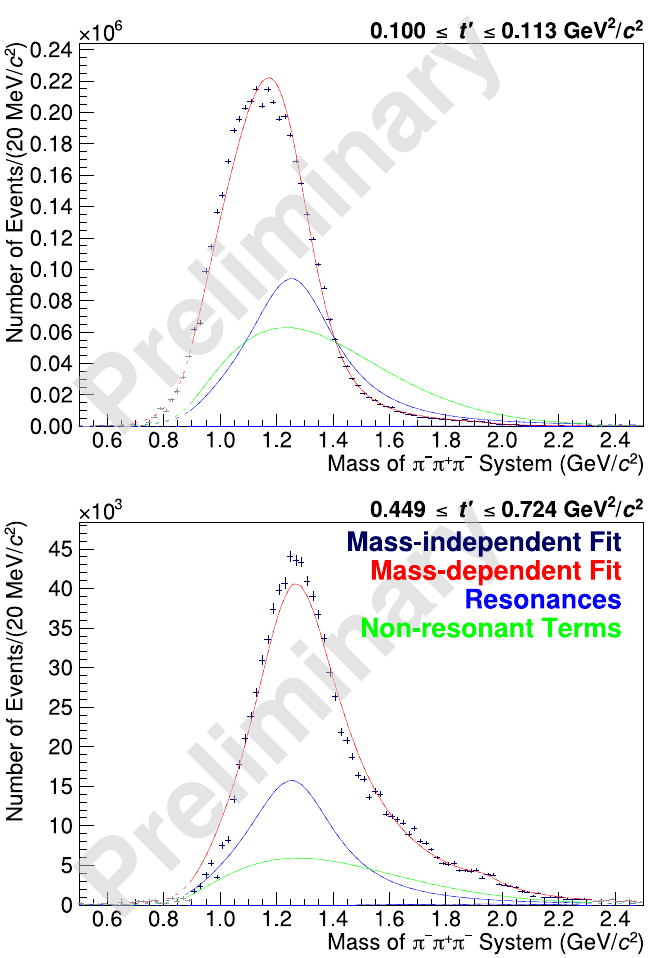}
  \includegraphics[width=0.32\columnwidth]{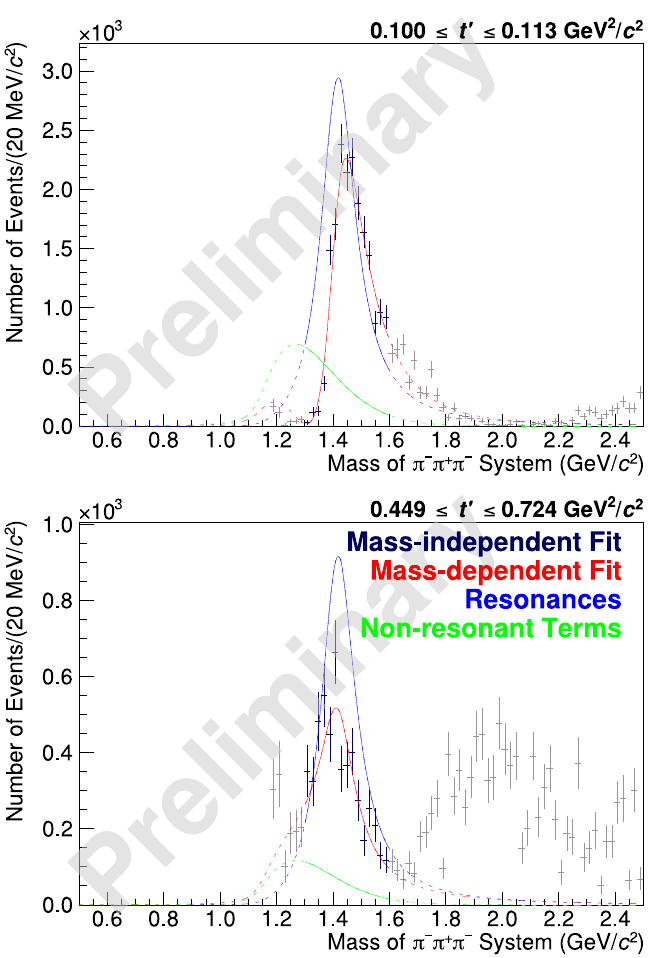}\\
  \caption{Intensities of (left) $1^{++}0^+\rho\pi\,S$ and
    (right) $1^{++}0^+f_0(980)\pi\,P$ waves, for (top) the smallest
    and (bottom) the 2nd highest $t'$-bin. The curves indicate the
    results of the fit to the spin-density matrix: (red) total
    intensity, (blue) resonance contribution, (green) background.} 
  \label{fig:pi-pi-pi+.a1}
\end{figure}

Some of the waves under study exhibit a strong $t'$-dependence, as can
be seen from Fig.~\ref{fig:pi-pi-pi+.a1} for the $1^{++}$ waves. In
the $1^{++}0^+\,\rho\pi\,S$ wave (left), the broad structure is
composed of the genuine $a_1(1260)$ resonance (which does not depend
on $t'$), and a background interfering destructively (upper plot: low
$t'$) or constructively (lower plot: high $t'$) at higher masses,
causing the resulting signal to change as a function of $t'$. The 
2D-analysis for the first time allows us to disentangle background and
resonance components. 
A similar picture is observed for the $a_1(1420)$, shown in the
right column of Fig.~\ref{fig:pi-pi-pi+.a1}. Note that the fit to the
spin-density matrices was
performed only in the mass range indicated by solid lines. 

\subsection{Model-independent Analysis of $\pi\pi$ $S$-Wave}
\label{sec:diff.deiso}
\begin{figure}[tbp]
  \centering
  \includegraphics[width=0.32\columnwidth]{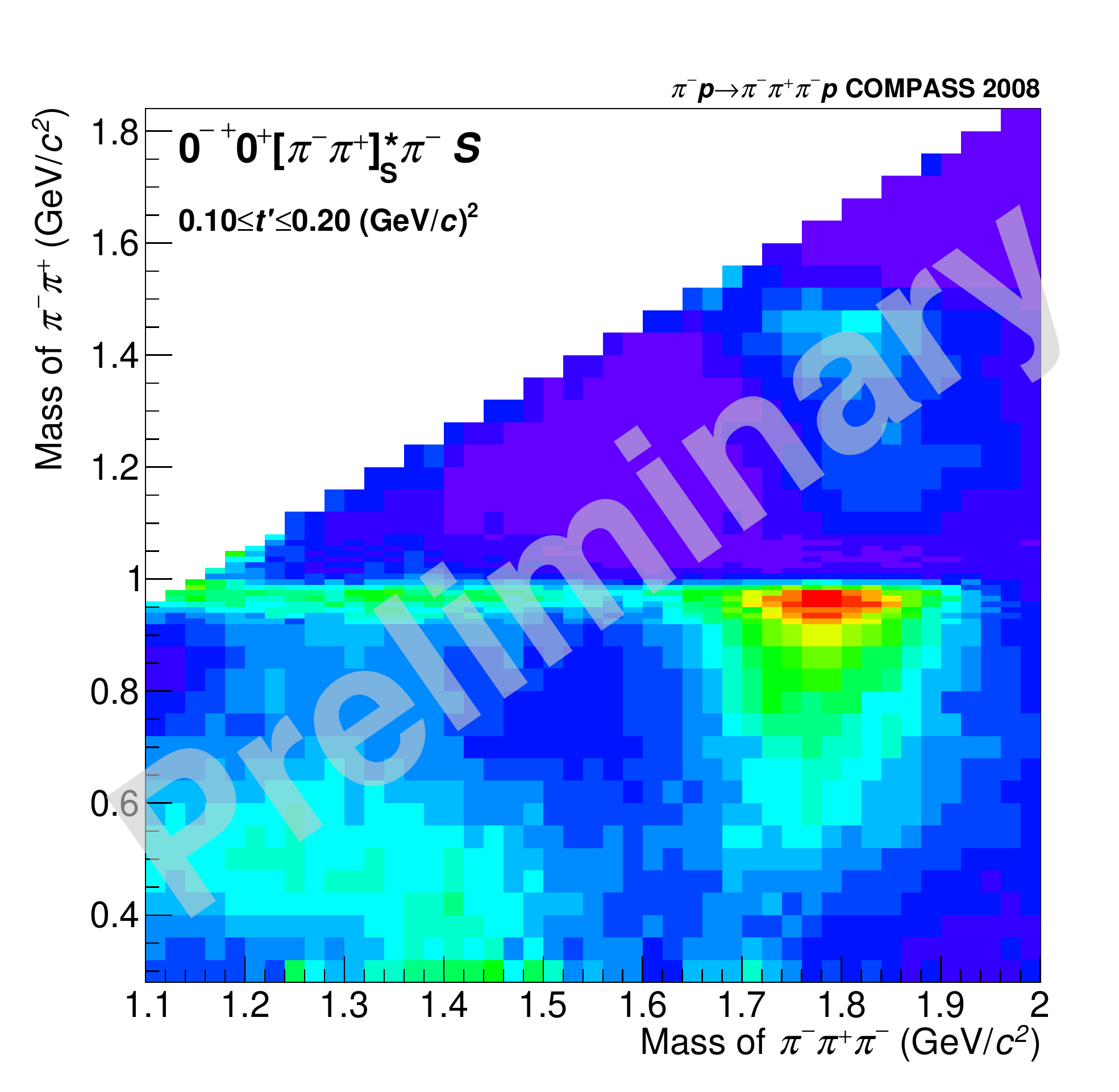}
  \hfill
  \includegraphics[width=0.32\columnwidth]{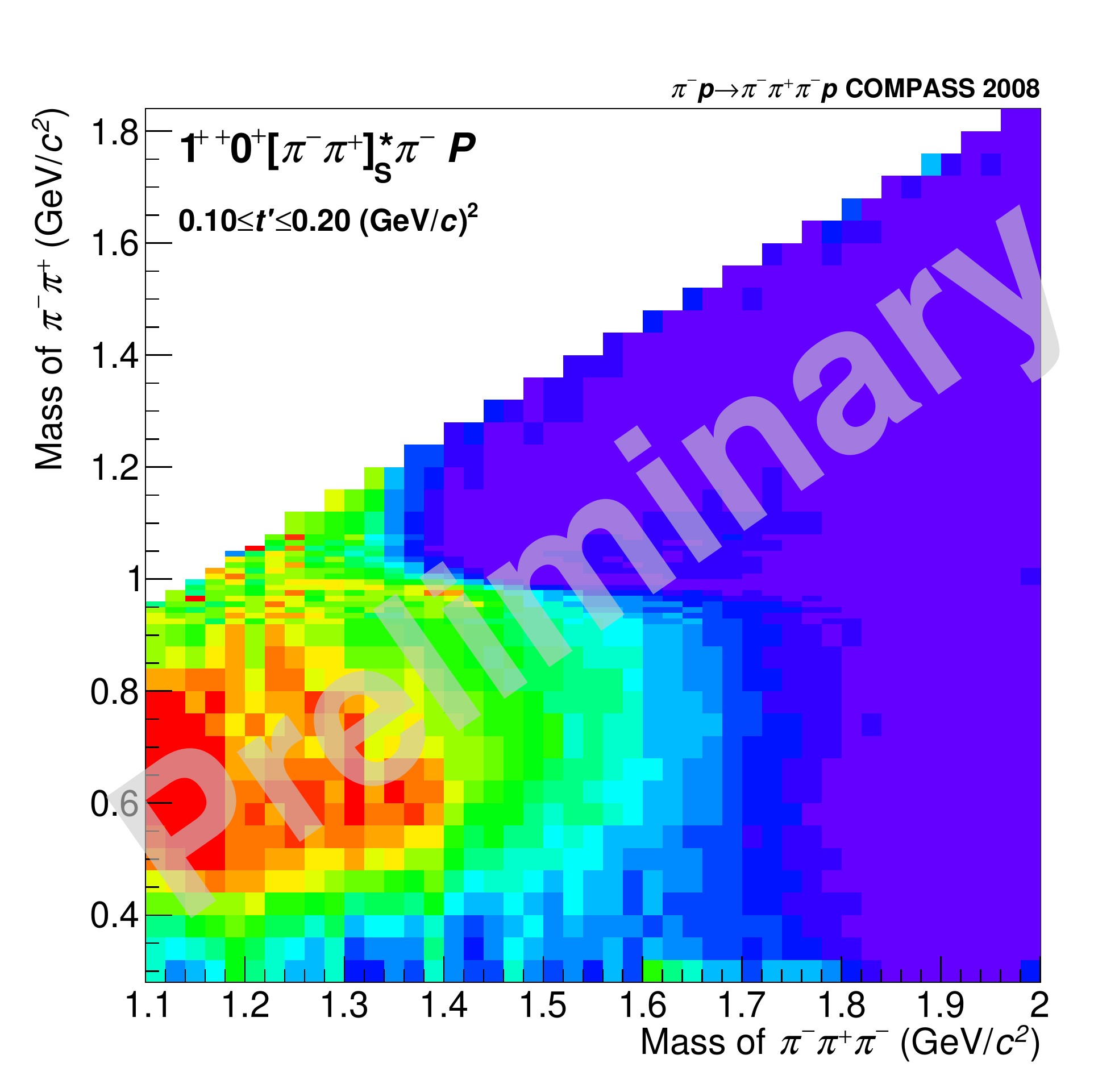}
  \hfill
  \includegraphics[width=0.32\columnwidth]{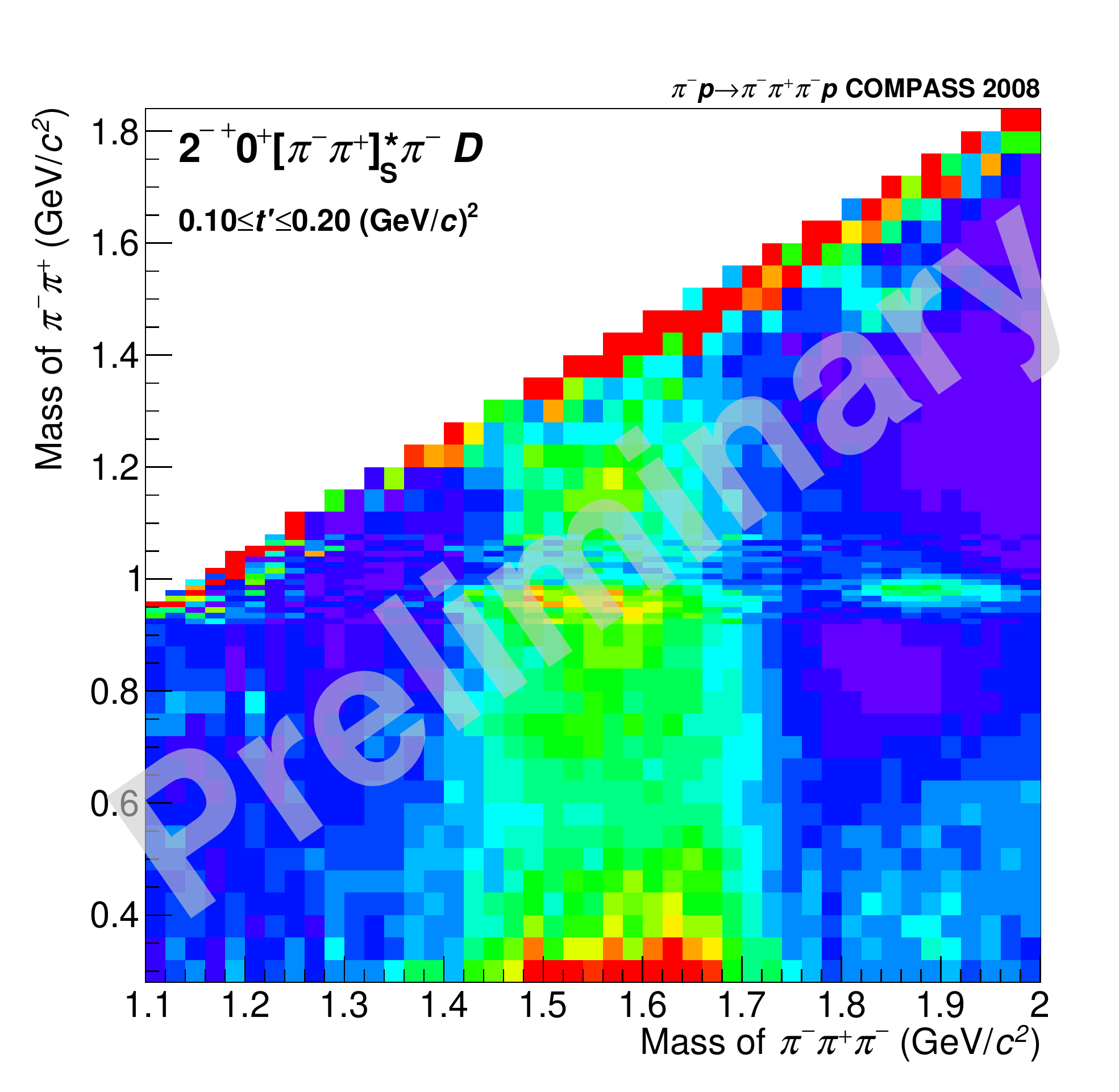}
  \includegraphics[width=0.32\columnwidth]{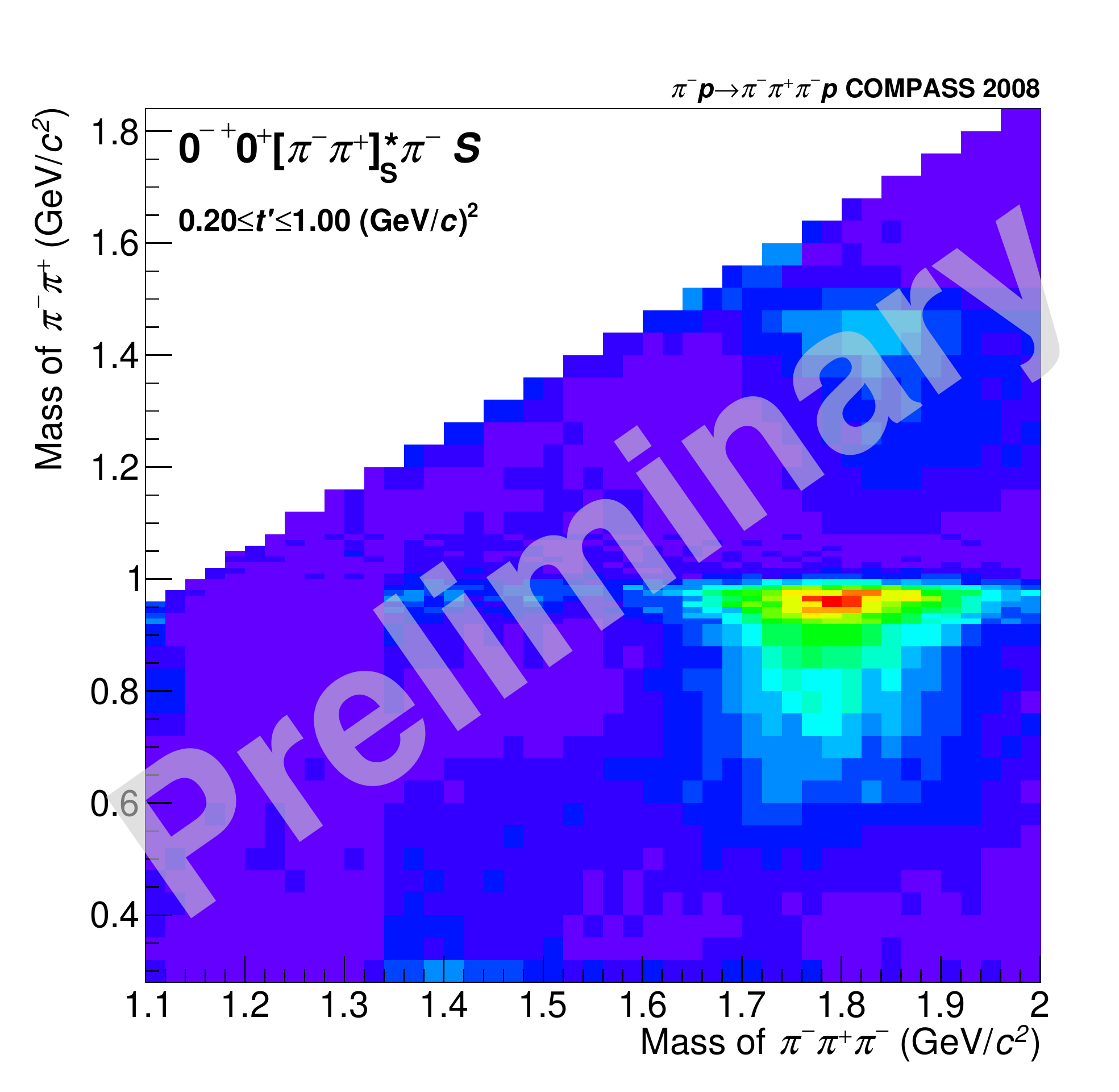}
  \hfill
  \includegraphics[width=0.32\columnwidth]{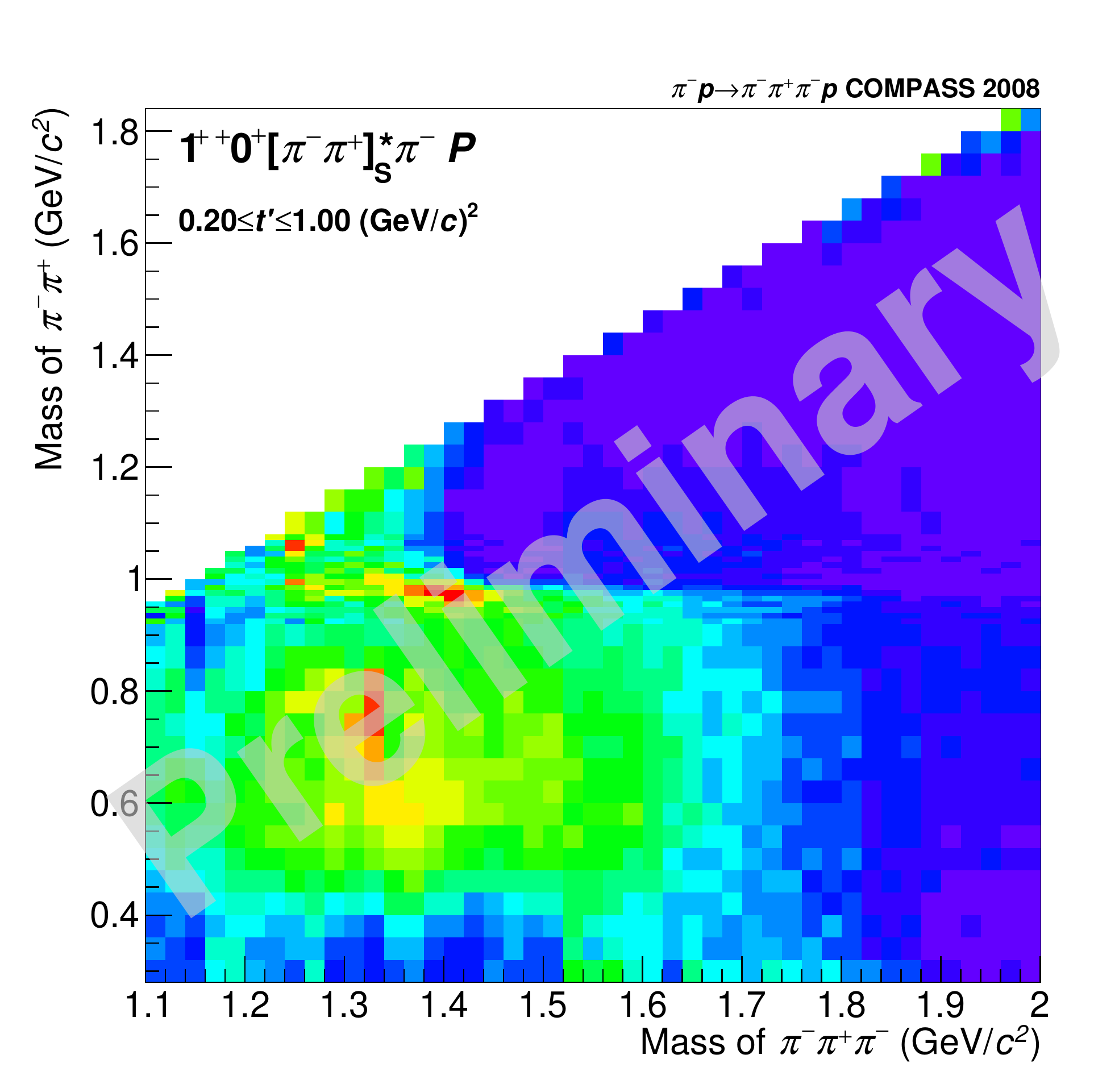}
  \hfill
  \includegraphics[width=0.32\columnwidth]{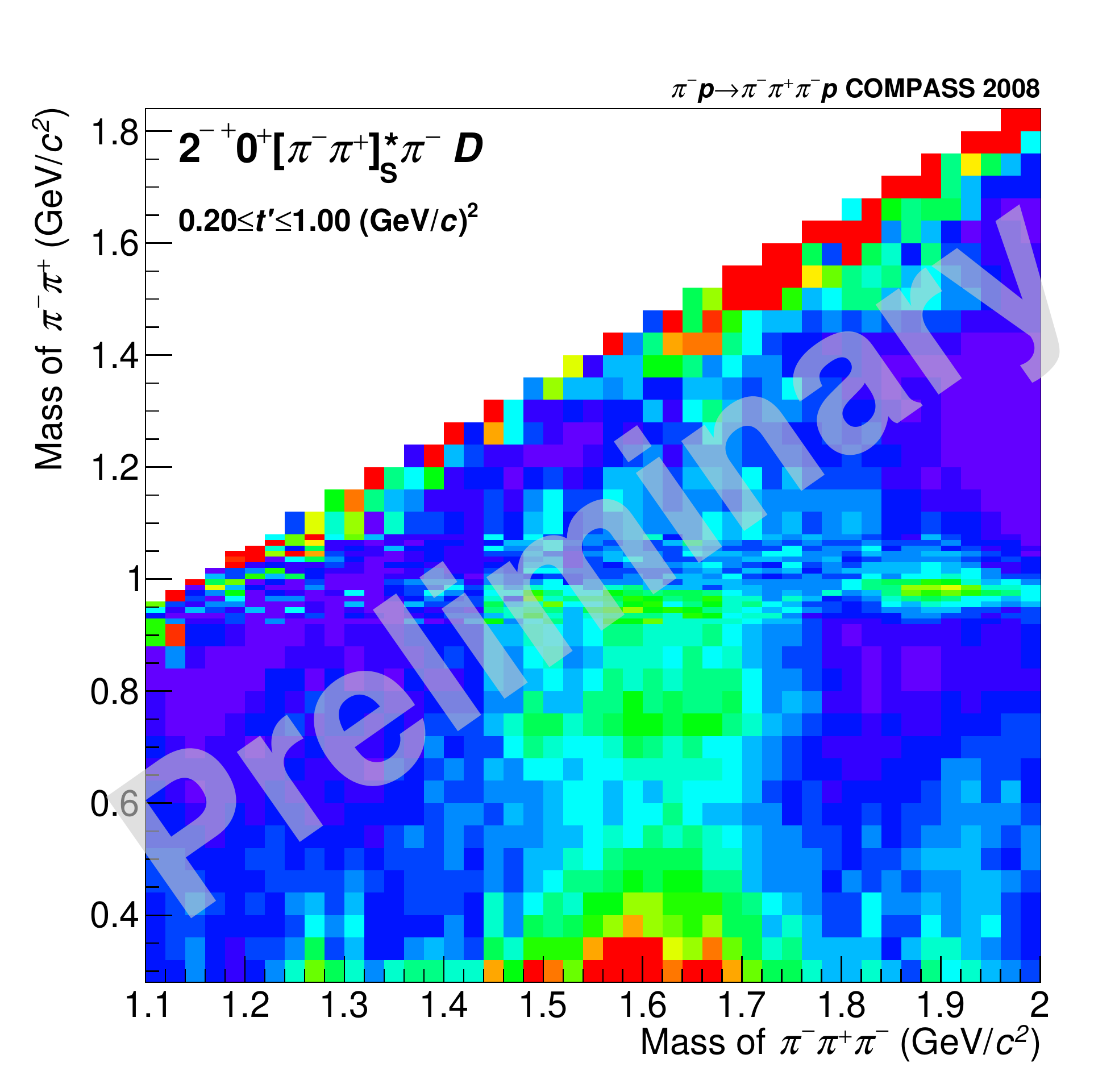}
  \caption{Intensity of waves as a function of 3-body invariant mass $m_{3\pi}$ and
    2-body mass $m_{2\pi}$ from fit with free $0^{++}$ 
    isobar amplitude, (left) $0^{-+}0^+(\pi\pi)_S^\ast\pi\,S$ wave, (center)
    $1^{++}0^+(\pi\pi)_S^\ast\pi\,P$ wave, (right) $2^{-+}0^+(\pi\pi)_S^\ast\pi\,D$
    wave. (Top row) $0.1\leq t'\leq 0.2\,\GeV^2/c^2$, (bottom row)
    $0.2\leq t'\leq 1.0\,\GeV^2/c^2$.} 
  \label{fig:pi-pi-pi+.m3pi-vs-m2pi}
\end{figure}

In order to study the dependence 
of our results on
the parameterization of isobars, especially the ones with
$J_\mathrm{iso}^{PC}=0^{++}$, 
we developed a method to determine the isobar dynamics directly from
the data, and thus to obtain a model-independent 
amplitude for these isobars. 
In this method the fixed 2-body amplitude for
$J_\mathrm{iso}^{PC}=0^{++}$,  
which includes the parameterization by \cite{Au:1986vs} of the broad
$f_0(500)$ and $f_0(1370)$ as well as a Flatt\'{e} ansatz for 
the narrow $f_0(980)$ and a Breit-Wigner for the $f_0(1500)$, is
replaced for three waves, $0^{-+}0^+\,(\pi\pi)_S^\ast\pi\,S$,
$1^{++}0^+\,(\pi\pi)_S^\ast\pi\,P$, and $2^{-+}0^+\,(\pi\pi)_S^\ast\pi\,D$,   
by a set of free complex parameters in 2-body mass bins of
$40\,\MeV/c^2$ width ($10\,\MeV/c^2$ around the $f_0(980)$),
denoted $(\pi\pi)_S^\ast$. 
There is
no separation into  
different $0^{++}$ isobars. Since the
number of free parameters is much larger in this case (one complex
parameter for each $m_{2\pi}$ mass bin per wave instead of one complex
parameter for each wave), the fit is
performed in two $t'$ bins only. Figure~\ref{fig:pi-pi-pi+.m3pi-vs-m2pi}
depicts the result of this fit for the three waves mentioned above
(columns) and for low $t'$ (upper row) and high $t'$ (lower row). 
A clear correlation of the intensity between 3-body and 2-body mass is
visible for all three waves. The left column shows the strong coupling of the
$\pi(1800)$ to the $f_0(980)$ and $f_0(1500)$. The middle column
proves the coupling of the narrow peak of the
$1^{++}0^+\,(\pi\pi)_S^\ast\pi\,P$ wave at masses around
$1.4\,\GeV/c^2$ to $f_0(980)$, which confirms its observation with the
fixed isobar parameterization. In the right column the $\pi_2(1880)$
decaying to $f_0(980)$ and $f_0(1500)$ is clearly visible. The
resonant nature of these $2\pi$ states is also visible from the full
complex amplitudes, which exhibit clear circular motions in the
respective Argand diagrams. 
There is a striking dependence of these distributions on $t'$ (upper
vs. lower row), especially in the $1^{-+}0^+\,(\pi\pi)_S^\ast\pi\,P$
wave (center). The very broad structure around 
$m_{3\pi}=1.3\,\GeV/c^2$ in the $0^{-+}0^+(\pi\pi)_S\pi\,S$, commonly
referred to as the $\pi(1300)$, 
exhibits no phase motion and a steeply falling $t'$ dependence, which
suggests it to be generated by non-resonant processes. 

\section{Photoproduction Reactions}
\label{sec:primakoff}
We now turn to reactions of the incoming pion with a heavy target
nucleus at very small 
values of 4-momentum transfer $t' \lsim \EE*{-3}\,\GeV^2/c^2$.
In this kinematic 
domain, the exchange of quasireal photons 
(Primakoff reaction) dominates over the strong interaction. The
observation of these reactions gives access to the low-energy
structure of 
hadrons, e.g.\ their polarizabilities, 
which are covered elsewhere in these proceedings, 
and to the coupling of hadronic 
resonances to photons, which is treated here.  

\begin{figure}[tbp]
  \centering
  \includegraphics[width=0.40\columnwidth]{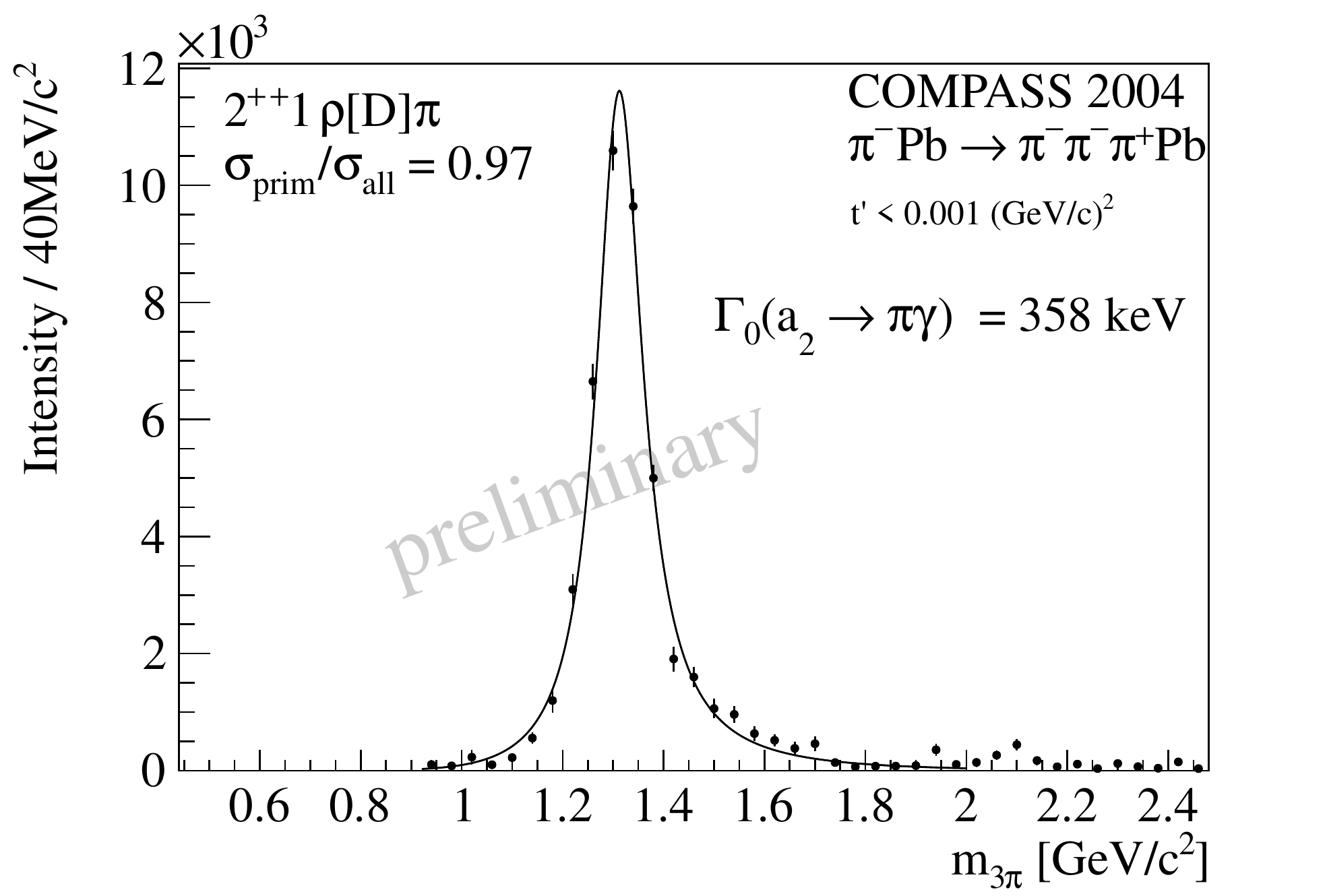}
  \includegraphics[width=0.40\columnwidth]{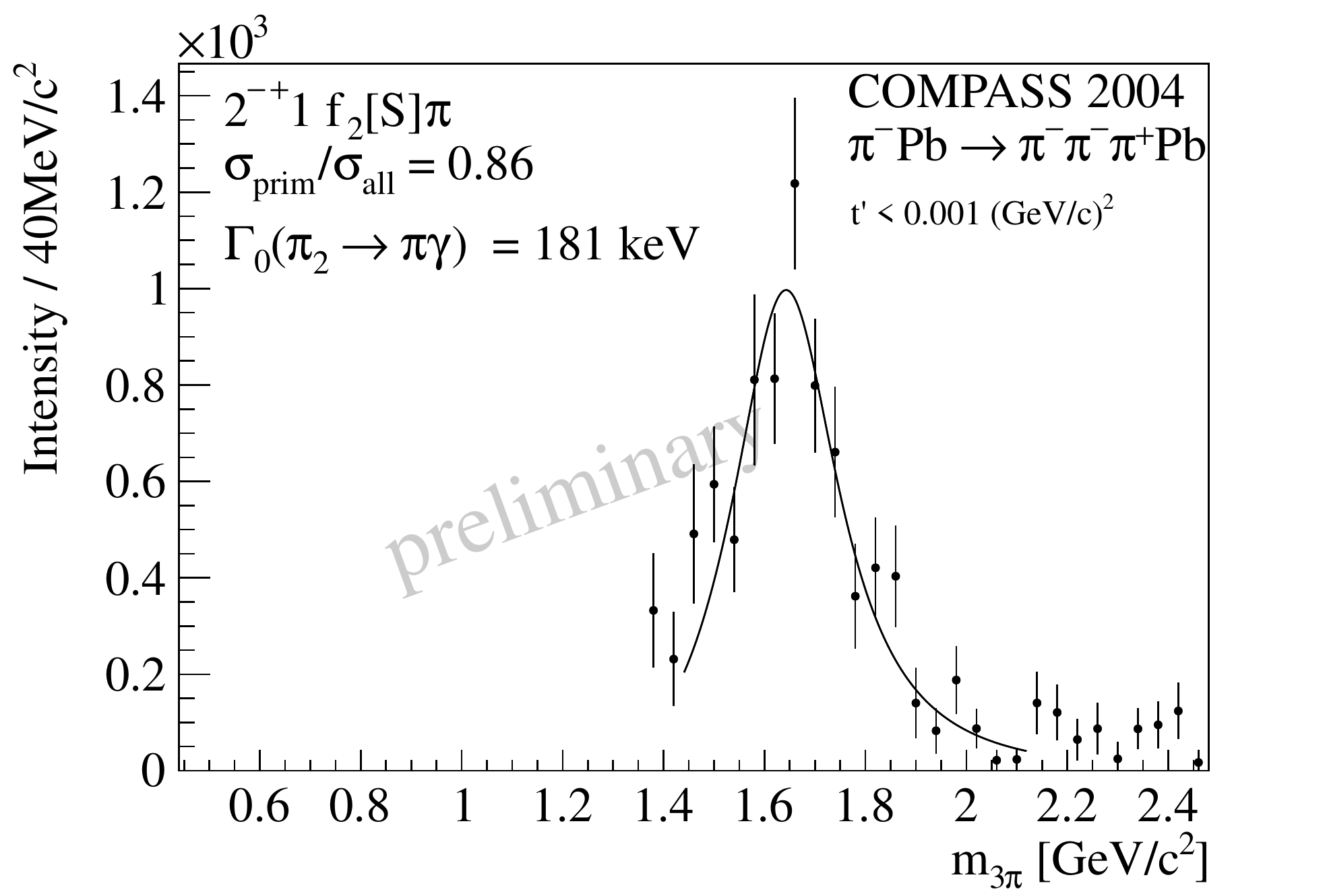}
  \caption{Intensities of (left) $2^{++}1\rho\pi\,D$ wave and (right)
    $2^{-+}1 f_2\pi\,S$ wave in the Primakoff region
    $t'<0.001\,\GeV^2/c^2$, together with the Breit-Wigner fits used
    to determine the radiative coupling.}  
  \label{fig:pi-pi-pi+.rad_width}
\end{figure}

We use about $\EE*{6}$ exclusive $\pi^-\pi^-\pi^+$ events produced on a Pb
target with values of $t'<0.001\,\GeV^2/c^2$ to perform a PWA similar
to the one described in Sec.~\ref{sec:diff.indep}. 
At these values of 4-momentum transfer 
both diffraction and
photoproduction contribute to
the production of a resonance $X$. 
With a dependence of the diffractive cross section on $t'$ as 
$t'^{M}\exp{(-b_\mathrm{diff}t')}$, however, only states with
spin projection $M=0$ are produced as $t'\rightarrow 0$, while states
with $M=1$ are dominantly produced by Primakoff reactions, which
depend on $t'$ as $\exp{(-b_\mathrm{prim}t')}$. 
The PWA then allows us to single out contributions of waves
with $M=1$, providing very clean access to quasireal
photoproduction and thus to the
radiative coupling of resonances decaying into $3\pi$. In the PWA 
clean signals in the $2^{++}1^+\,\rho\pi\,D$ wave, corresponding to
$\pi\gamma\rightarrow a_2(1320)$, and in the $2^{-+}1^+\,f_2\pi\,S$
wave, corresponding to $\pi\gamma\rightarrow \pi_2(1670)$ are
observed, as shown in Fig.~\ref{fig:pi-pi-pi+.rad_width}.  


The radiative width $\Gamma(X\rightarrow \pi\gamma)$ is extracted from
the measured Primakoff cross 
section $\sigma_\mathrm{prim}$ by relating it to the integral
of the double differential cross section for Primakoff production of a
wide Breit-Wigner resonance (given e.g.\ in \cite{Molchanov:2001qk}),
calculated in the 
Weizs\"acker-Williams equivalent-photon approximation (EPA), over the
experimentally observed mass and $t'$ range, 
\begin{equation}
  \label{eq:rad_width}
  \sigma_\mathrm{prim}=\int_{m_1}^{m_2} \int_0^{0.001}
  \diff{m}\diff{t'}
  \frac{\diffsq{\sigma}}{\diff{m}\diff{t'}}
  =\Gamma_0(X\rightarrow\pi\gamma)\cdot C_X\quad.
\end{equation}
Here, $C_X$ is a normalization constant
which includes the total width of the resonance under study, the
partial width in the final 
state, and several kinematic factors related to the dynamics of the
initial and final states, and the EPA. The curves in
Fig.~\ref{fig:pi-pi-pi+.rad_width} show the result of a fit of the
mass-dependent differential cross section in Eq.~(\ref{eq:rad_width})
integrated over the relevant $t'$ range to the data points extracted
from the PWA in mass bins. The integral of the fitted curve gives
$\sigma_\mathrm{all}$, from which the small diffractive contribution,
estimated from a fit of the $t'$-dependence of the intensities, 
is subtracted incoherently to give $\sigma_\mathrm{prim}$. This is
justified due to the fact that the 
phase difference between photoproduction and diffractive production is
close to $90^\circ$ for the small values of $t'$ used in this
analysis. 

The COMPASS result for the radiative width of the
$a_2(1320)$ is $\Gamma_0(a_2(1320)\rightarrow\pi\gamma)=(358\pm 6\pm
42)\,\keV$, in fair agreement with calculations based on the vector
meson dominance model \cite{Rosner:1980ek} and a relativistic
quark model \cite{Aznauryan:1988zz}. For the $\pi_2(1670)$ the
experimental result is
$\Gamma_0(\pi_2(1670)\rightarrow\pi\gamma)=(181\pm 11\pm 
27)\,\keV \cdot \left(0.56/B\!R_{f_2\pi}\right)$, where the correction
factor indicates that the branching ratio
$B\!R_{f_2\pi}^\mathrm{PDG}=0.56$ taken from 
\cite{Beringer:2012zz} was used for the quoted
result, which may have to be modified when interferences between the
different $3\pi$ final states are taken into account. 
The systematic errors include contributions from radiative
corrections, the determination of the luminosity from the decay of
beam kaons, different PWA models, the subtraction of the diffractive
background, and the mass-dependent parameterization of the resonance
shape. The radiative
width of the $\pi_2(1670)$ quoted here is the first experimental
determination of this quantity, and differs significantly from the
theoretical prediction by \cite{Maeda:2013dka}. 

\section{Conclusions}
\label{sec:conclusions}
The COMPASS experiment has collected large data samples on diffractive
and photon-induced reactions of light mesons. For the largest data set
with $3\pi$ final
states a 2-dimensional PWA method in bins of
invariant mass and 4-momentum transfer was established, which allows
the disentanglement 
of resonant and non-resonant contributions and
thus a precise determination of the parameters for a number
of resonances decaying into $3\pi$, including radial excitations of
the $a_1(1260)$ and $a_2(1320)$. 
A new resonant structure with axial-vector quantum numbers is observed
in the $1^{++}0^+\,f_0(980)\,\pi\,P$ wave at a mass of
$1.42\,\GeV/c^2$, which we tentatively call $a_1(1420)$ with a
Breit-Wigner mass of $1412$-$1422\,\MeV/c^2$ and a rather narrow width
of $130$-$150\,\MeV/c^2$.  

In addition, a new model-independent method to determine the
amplitude of contributing isobars from the data, rather than using
fixed parameterizations, was developed. It was first applied to scalar 
isobars with $J_\mathrm{iso}^{PC}=0^{++}$ coupling to two pions in an
$S$ wave,
which play an important role in many decay channels. 
Strong coupling and isobar phase motions close to $180^\circ$,
corresponding to the 
$f_0(980)$ and $f_0(1500)$, are observed for the $\pi(1800)$ and the
$\pi_2(1880)$ mother states. 
The $a_1(1420)$ is also seen with this method coupling solely to
$f_0(980)\pi$, confirming the result with fixed scalar isobar
parameterizations.  
A strong dependence of the shape of the
scalar isobar amplitude on the mass of the mother state as well as on
$t'$ is observed. The new method allows us to reduce the
model-dependence of the PWA, albeit at a drastic
increase in the number of fit parameters. 

Using Primakoff reactions mediated by quasireal photons, the radiative
widths of the $a_2(1320)$ 
and, for the first time, the $\pi_2(1670)$ were determined to be
$\Gamma_0(a_2(132 0)\rightarrow\pi\gamma)=(358\pm 6\pm 42)\,\keV$ and 
$\Gamma_0(\pi_2(1670)\rightarrow\pi\gamma)=(181\pm 11\pm
27)\,\keV \cdot \left(0.56/B\!R_{f_2\pi}\right)$, respectively. 

\bibliographystyle{prsty}
\bibliography{hadron,compass,panda,detectors}

\end{document}